\documentclass[12pt,article]{emulateapj}

\shorttitle{Keck/OSIRIS IFU Detection of z$\sim 3$ Galaxy}
\shortauthors{H. M. Christenson \& R. A. Jorgenson}

\usepackage{graphicx}
\usepackage{amssymb}
\usepackage{amsmath}
\usepackage{cleveref}

\usepackage{natbib}
\begin{document}

\def\intl{\int\limits}
\def\nstat{$\approx $}
\def\perd{\;\;\; .}
\def\cmma{\;\;\; ,}
\def\ltk{\left [ \,}
\def\ltp{\left ( \,}
\def\ltb{\left \{ \,}
\def\rtk{\, \right  ] }
\def\rtp{\, \right  ) }
\def\rtb{\, \right \} }
\def\jnu{$J_{\nu}$}
\def\jnuphot{$J_{\nu}^{phot}$}
\def\jnuciistar{$J_{\nu}^{\rm CII^{*}}$}
\def\junit{ergs cm$^{-2}$ s$^{-1}$ Hz$^{-1}$ sr$^{-1}$}
\def\jnutot{$J_{\nu}$$^{total}$}
\def\jnubkd{$J_{\nu}$$^{Bkd}$}
\def\jnuloc{$J_{\nu}$$^{local}$}
\def\jnulw{$J_{\nu}$$^{LW}$}
\def\jnutotciistr{$J_{\nu }$$^{total, C\,II^*}$}
\def\jnulocciistr{$J_{\nu } $$^{local, C\,II^*}$}
\def\jnutotci{$J_{\nu }$$^{total, C\, I}$}
\def\jnulocci{$J_{\nu }$$^{local, C\, I}$}
\def\jnutothtwo{$J_{\nu }$$^{total, H_2}$}
\def\jnulochtwo{$J_{\nu }$$^{local, H_2}$}
\newcommand{\snrlim}{SNR$_{lim}$}
\newcommand{\nhi}{$N_{\rm HI}$}
\newcommand{\mnhi}{N_{\rm HI}}
\newcommand{\flls}{f_{\rm HI}^{\rm LLS}}
\newcommand{\fdla}{f_{\rm HI}^{\rm DLA}}
\newcommand{\llls}{$\ell_{\rm LLS}$}
\newcommand{\ldla}{\ell_{\rm DLA}}
\newcommand{\fnhi}{$f_{\rm HI}(N,X)$}
\newcommand{\mfnhi}{f_{\rm HI}(N,X)}
\newcommand{\Nth}{2 \sci{20} \cm{-2}}
\newcommand{\taux}{$d\tau/dX$}
\newcommand{\gz}{$g(z)$}
\newcommand{\nz}{$n(z)$}
\newcommand{\nx}{$n(X)$}
\newcommand{\omg}{$\Omega_g$}
\newcommand{\ostr}{$\Omega_*$}
\newcommand{\momg}{\Omega_g}
\newcommand{\olls}{$\Omega_g^{\rm LLS}$}
\newcommand{\odla}{$\Omega_g^{\rm DLA}$}
\newcommand{\oneut}{$\Omega_g^{\rm Neut}$}	
\newcommand{\ohi}{$\Omega_g^{\rm HI}$}
\newcommand{\olwz}{$\Omega_g^{\rm 21cm}$}
\newcommand{\ndla}{71}
\newcommand{\cmk}{cm$^{-3}$ K }
\newcommand{\kms}{{km\ s$^{-1}$}}
\newcommand{\flux}{ergs\ s$^{-1}$\ cm$^{-2}$}
\newcommand{\lum}{ergs\ s$^{-1}$}
\newcommand{\lya}{Ly$\alpha$}
\newcommand{\ci}{C\,I}
\newcommand{\cistr}{C\,I$^{*}$}
\newcommand{\mcistr}{C\,I^{*}}
\newcommand{\cistrstr}{C\,I$^{**}$}
\newcommand{\mcistrstr}{C\,I^{**}}
\newcommand{\citot}{(C\,I)$_{tot}$}
\newcommand{\mcitot}{(C\,I)_{tot}}
\newcommand{\cli}{Cl\,I}
\newcommand{\clii}{Cl\,II}
\newcommand{\cii}{C\,II}
\newcommand{\ciistr}{C\,II$^*$}
\newcommand{\dla}{DLA}
\newcommand{\dlas}{DLAs}
\newcommand{\htwo}{H$_{\rm 2}$}
\newcommand{\he}{He\, I}
\newcommand{\sii}{Si\,II}
\newcommand{\siistr}{Si\,II$^{*}$}
\newcommand{\hi}{H\, I}
\newcommand{\ctwo}{C\,II}
\newcommand{\ewsitwo}{$W_{\lambda 1526}$}
\newcommand{\ewciv}{$W_{\lambda 1548}$}
\newcommand{\feii}{Fe\,II}

\newcommand{\halpha}{H$\alpha $}
\newcommand{\hb}{H$\beta$}
\newcommand{\hbeta}{H$\beta\lambda$4861}
\newcommand{\ha}{H$\alpha $}
\newcommand{\hamath}{H\alpha }
\newcommand{\oiii}{[O\,III]}
\newcommand{\nii}{[N\,II]}
\newcommand{\oiiia}{[O\,III]$\lambda$5007}
\newcommand{\oiiib}{[O\,III]$\lambda$4959}

\newcommand{\mydla}{\dla 2233$+$131}

\newcommand{\myzabs}{3.153}
\newcommand{\hireszabs}{3.14927}

\newcommand{\gasfrac}{0.6}

\newcommand{\oiiiazabs}{3.1508} 
\newcommand{\oiiiazabsu}{3.1508 \pm 0.0001}
\newcommand{\oiiiavel}{110}
\newcommand{\oiiiaflux}{(2.4 \pm 0.5) \times10^{-17}}
\newcommand{\oiiialum}{(2.2 \pm 0.5) \times10^{42}}
\newcommand{\oiiiafwhm}{99 \pm 1.6}
\newcommand{\oiiiafwhminit}{141\pm1}
\newcommand{\oiiiamassb}{5.9\times10^9}
\newcommand{\oiiiamass}{3.1\times10^9} 
\newcommand{\sfra}{13.6 \pm 7.1} 
\newcommand{\sfrb}{7.1 \pm 1.6} 
\newcommand{\sfrc}{10.3 \pm 2.5} 
\newcommand{\oiiiasig}{13.2} 
\newcommand{\oiiiasignet}{42.0} 
\newcommand{\oiiiacen}{2078.8 \pm 0.1} 
\newcommand{\totgasa}{6.0 \times 10^9}
\newcommand{\totgasb}{3.8 \times 10^9}
\newcommand{\totgasc}{4.9\times 10 ^9}
\newcommand{\totgas}{(3.8 $-$ 6.0) \times10^9} 

\newcommand{\oiiibzabs}{3.1508} 
\newcommand{\oiiibzabsu}{3.1508 \pm 0.0001} 
\newcommand{\oiiibvel}{108}
\newcommand{\oiiibflux}{(2.0 \pm 7.5) \times10^{-18}}
\newcommand{\oiiiblum}{(1.8 \pm 6.8) \times10^{41}}
\newcommand{\oiiibfwhm}{101 \pm 1.6}
\newcommand{\oiiibfwhminit}{143\pm1}
\newcommand{\oiiibcen}{2058.9 \pm 0.1} 

\newcommand{\oiiibsig}{1.8} 

\newcommand{\hbetazabs}{3.1518} 
\newcommand{\hbetazabsu}{3.1518 \pm 0.01}
\newcommand{\hbetasig}{3.1}
\newcommand{\hbetavel}{186}
\newcommand{\hbetaflux}{(6.8\pm 3.6) \times10^{-18}}
\newcommand{\hbetalum}{(6.2 \pm 3.2) \times10^{41}}
\newcommand{\hbetafwhm}{NaN} 
\newcommand{\hbetafwhminit}{74\pm 2} 

\newcommand{\specres}{3000} 
\newcommand{\specresKMS}{101}
\newcommand{\oversample}{0.05}
\newcommand{\kernelFWHM}{4}
\newcommand{\quasarsep}{2.3}
\newcommand{\quasardist}{17.8}
\newcommand{\fwhmpsf}{0.15}
\newcommand{\sigins}{101}
\newcommand{\kpcscale}{7.739}
\newcommand{\spaxelsize}{0.0025}

\newcommand{\zabs}{$z_{abs}$}
\newcommand{\zem}{$z_{em}$}
\newcommand{\sfr}{M$_{\odot}$ yr$^{-1}$}

\def\lc{${\ell}_{c}$}
\def\lcunit{ergs s$^{-1}$ H$^{-1}$}


\title{Keck/OSIRIS IFU\altaffilmark{*} detection of a z$\sim 3$ damped Lyman alpha host galaxy} 

\author{Holly M. Christenson \altaffilmark{1,2,3} \& Regina A. Jorgenson \altaffilmark{3}}

\altaffiltext{1}{University of California, Riverside, 900 University Avenue, Riverside, CA 92521; holly.christenson@email.ucr.edu}
\altaffiltext{2}{Western Washington University, 516 High Street, Bellingham, WA 98225}
\altaffiltext{3}{Maria Mitchell Observatory, 4 Vestal Street, Nantucket, MA 02554; rjorgenson@mariamitchell.org}
\altaffiltext{*}{The data presented herein were obtained at the W. M. Keck Observatory, which is operated as a scientific partnership among the California Institute of Technology, the University of California and the National Aeronautics and Space Administration. The Observatory was made possible by the generous financial support of the W. M. Keck Foundation.}

\begin{abstract}
We present Keck/OSIRIS infrared IFU observations of the $z = $ \myzabs\ sub-DLA \mydla, previously detected in absorption to a background quasar and studied with single slit spectroscopy and PMAS integral field spectroscopy (IFU).  We used the Laser Guide Star Adaptive Optics (LGSAO) and OSIRIS IFU to reduce the point-spread function of the background quasar to FWHM$\sim$\fwhmpsf\arcsec\ and marginally resolve extended, foreground \dla\ emission. We detect \oiiia\ emission with a flux F$^{\oiii \lambda5007}$ = $\oiiiaflux$ \flux, as well as unresolved \oiiib\ and \hbeta\ emission. Using a composite spectrum over the emission region, we measure dynamical mass $\sim$ $\oiiiamass$ M$_{\odot}$. We make several estimates of star formation rate using \oiiia\ and \hbeta\ emission, and measure a star formation rate of $\sim$ $7.1- 13.6$ \sfr. We map \oiiia\ and \hbeta\ emission and the corresponding velocity fields to search for signs of kinematic structure. These maps allow for a more detailed kinematic analysis than previously possible for this galaxy.  While some regions show slightly red and blue-shifted emission indicative of potential edge-on disk rotation, the data are insufficient to support this interpretation.

\end{abstract}


\section{Introduction} 

 Studying galaxies at high redshift has always been challenging given the great distances and natural dimming of galaxies' light.  While some galaxies have now been detected in emission at extremely high redshifts (for example, ~\citealt{genzel06}), these studies can only probe the highest luminosity, most rapidly star-forming galaxies. These galaxies are likely not representative of the galaxy population as a whole; the bulk of the galaxy population is forming stars at a much more modest rate, and thus cannot be probed directly by extremely high-redshift studies.    

To have a complete picture of galaxy formation and evolution, a more representative sample of the galaxy population is needed. Damped Lyman alpha systems (\dlas) are an ideal candidate for probing the typical galaxy population. \dlas\ contain the bulk of neutral hydrogen in the z $\sim$ 2 -- 5 universe, the raw material for star formation, and are thought to be the progenitors of modern spiral galaxies ~\citep{wolfe2005}. \dlas\ are defined by their column density in neutral hydrogen, and are detected in absorption to background light sources, typically quasars. The first large survey of \dlas\ was completed by ~\cite{wolfe86} and since then, thousands have been studied in absorption (see ~\citealt{blanton17} for an overview). Absorption studies provide detailed information about redshift, neutral hydrogen column density, and metallicity. However, nearly all absorption line studies are inherently limited in scope because they probe only a single line of sight through a given galaxy.

In an effort to remediate the problems inherent with absorption-line-only studies, practitioners have been trying to detect DLA galaxies directly in emission for many years.  This is a tremendously difficult task, given the difficulty of detecting relatively faint foreground emission towards a much brighter background quasar (i.e. ~\citealt{lowenthal95, bunker99, kulkarni00, kulkarni06, christensen09}).  To date, only $\sim $ 15 $z > $ 2 \dlas\ have been detected in emission (see ~\citealt{krogager17} for a summary).  Most of these targets were detected in single slit observations, requiring fortuitous slit placement and providing limited information on the total fluxes, star formation rates (SFR), and kinematics of the emission. 

Technological advances made in the past two decades, in form of Integral Field Units (IFUs) on 10-meter class telescopes assisted by laser guide star adaptive optics (LGSAO) systems, are now enabling the direct detection of these elusive galaxies.  While progress is slow, given the relatively small fluxes and correspondingly long exposures required (see for example ~\citealt{jorgenson14}), this is an important task not only for gaining a complete understanding of the bulk of the high$-z$ galaxy population, but also in preparation for similar studies that will be made possible (and more efficient) with space telescopes such as James Webb Space Telescope.

In this paper, we present observations using the Keck/OSIRIS IFU ~\citep{larkin06} with LGSAO to target the super Lyman limit system, or sub-\dla , \mydla , with N(HI) = 1 $\times$ 10$^{20}$ cm$^{-2}$
. We note that this column density is just under the DLA threshold of N(HI) = 2 $\times$ 10$^{20}$ cm$^{-2}$, although \mydla\ is most often referred to as a \dla\ in the literature. We will do so here for consistency. 

\mydla\ is a well studied system - first discovered by ~\cite{sargent1989}, who reported the \dla\ redshift of $z = $ 3.153. \mydla\ was first detected in emission by ~\cite{steidel95} who performed deep broadband imaging of a sample of Lyman limit systems in order to search for the counterparts of known Lyman Limit systems.  ~\cite{steidel95} report the detection of R band stellar continuum from a system located 2.9\arcsec\ from the background quasar Q2233$+$131.  Several more reports of emission followed, including ~\cite{djorgovski96}, who report the detection of \mydla\ in both stellar continuum and in \lya\ line emission located 2.3\arcsec\ away at PA = 159$^{\circ}$ from the quasar line of sight.  

~\cite{warren01} and ~\cite{moller02} report HST/NICMOS imaging of \mydla\ with magnitude H$_{AB}$=25.05 located 2.78\arcsec\ away from the background quasar at PA=158.6$^{\circ}$.

Finally, ~\cite{christensen04} used the Potsdam Multi Aperture Spectrophotometer (PMAS) to measure extended \lya\  flux of \mydla , with a total flux of (2.8 $\pm$ 0.3) $\times$ 10$^{-16}$ erg cm$^{-2}$ s$^{-1}$ measured over an area $3\times5$\arcsec.  ~\cite{christensen04} estimate the star formation rate (SFR) to be 19 $\pm$ 10 \sfr\ and conjecture that the extended flux may be powered by a star formation fueled outflow from the \dla\  galaxy. ~\cite{christensen07} perform higher spectral resolution PMAS observations resulting in \lya\ flux of (9.6 $\pm$ 2.5) $\times$ 10$^{-17}$ erg cm$^{-2}$ s$^{-1}$, consistent with the previous \cite{djorgovski96} measurement (see Table~\ref{tab:previous} for a summary of previous work).

 In this work, we map the flux and velocity field of \oiii\ and \hb\ emission from \mydla\ in order to look for kinematic signatures that may help reveal the underlying nature of the galaxy.  The paper is organized as follows:  We describe the observations and data reduction process in Section~\ref{sec:obs}.  In Section~\ref{sec:analysis} we discuss the details of the analysis of the final spectral cube.  We place these results in a larger context in Section~\ref{sec:results}, before summarizing in Section~\ref{sec:summary}.  Throughout the paper we assume a standard lambda cold dark matter ($\Lambda$CDM) cosmology based on the final nine-year {\it WMAP} results ~\citep{wmap} in which $H_0$ = 70.0 km s$^{-1} \mathrm{Mpc}^{-1}$, $\Omega _{m}$ = 0.279 and $\Omega _{\Lambda}$ = 0.721.

\section{Observations}~\label{sec:obs}

\begin{figure*}
\includegraphics[scale=0.4]{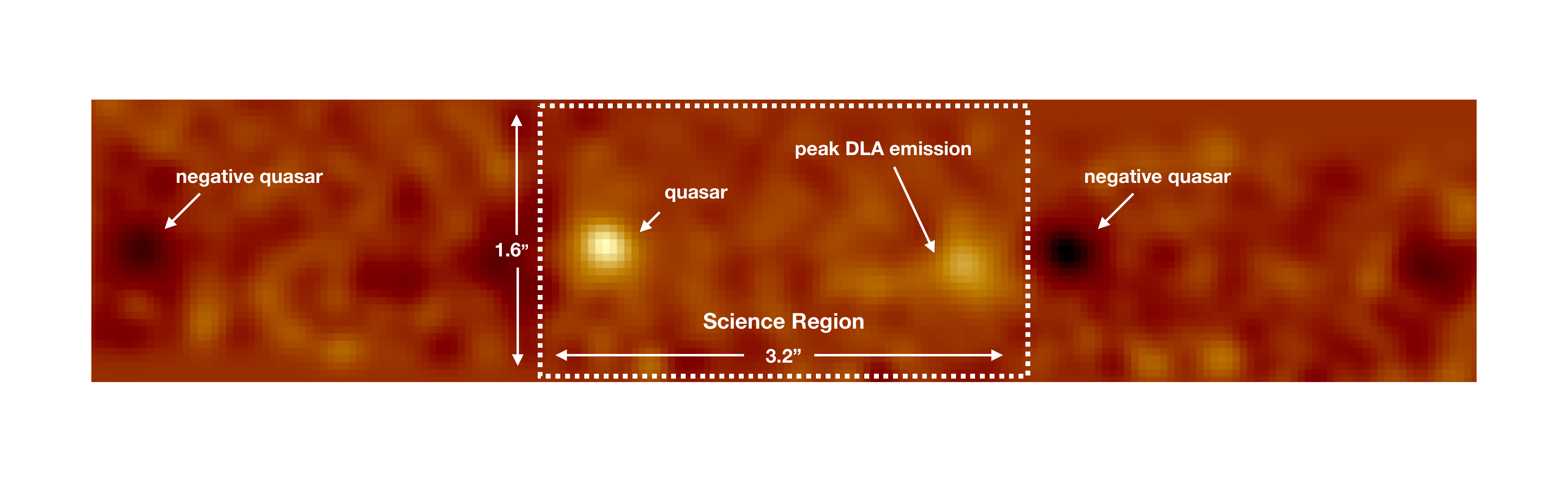}
\caption{Final spectral cube summed over redshifted \oiiia\ emission. The sky subtraction scheme creates two negative images of the quasar. The central,  usable science region, 3.2\arcsec$\times$1.6\arcsec\ in size,  is boxed.  This region remains positive after the background subtraction and encompasses both the positive image of the quasar and the peak of \dla\ emission. 
}
\label{fig:cube}
\end{figure*}

Observations were made with the Keck/OSIRIS ~\citep{larkin06} infrared integral field spectrograph and the Keck I LGSAO system on 2014 July 18, using the broadband Kbb filter that covers 1965 to 2381 nm. This wavelength range covers the redshifted wavelengths of \oiii\ and \hb\ emission from \mydla\ at $z \sim$ 3.15.  Unfortunately, \ha\ emission is redshifted out of the observable OSIRIS wavelength range. We used the 100 mas plate scale, giving a 1.6$\times$ 6.4\arcsec\ field of view. 
The average spectral resolution for OSIRIS in this configuration is R $\sim$ \specres\ (100 \kms), with small variations from spaxel to spaxel.

We aligned the detector at a position angle of PA = 158.45$^{\circ}$ East of North ~\citep{weatherley05}, in order to detect both the quasar and the peak of DLA emission in a single exposure.  To maximize on-target efficiency, we used an A-B observing pattern, shifting the quasar and DLA emission pair from the bottom half of the of field of view to the top half of the field of view for each exposure.  This observing pattern allowed us to use each alternate exposure as the sky-frame for the previous on-target frame.    
The field of view is effectively halved after sky background subtraction, but the on-source observing time is doubled, increasing the likelihood of detecting faint, extended emission. In total, we obtained 13 $\times$ 900 second exposures, for a total on-source exposure time of 3.25 hours.

We chose a star within 50\arcsec\ with R$<$17 to make tip tilt corrections. Observations were made in good weather, with effective seeing, after LGSAO corrections, of 0.15\arcsec.

\subsection{Data Reduction and Flux Calibration}
Observations were reduced using a combination of the Keck/OSIRIS data reduction pipeline (DRP) and a set of custom IDL routines, similar to the procedures outlined in ~\cite{law07, law09} and ~\cite{jorgenson14}. The DRP was used for standard reduction and extraction of three-dimensional spectral cubes from the raw data. 

We then took the following additional steps:

\begin{enumerate}
    \item \textbf{Background sky subtraction.} We used the OSIRIS DRP background sky subtraction routine to perform the A-B subtraction as previously described.
    \item \textbf{Second pass sky subtraction.} To reduce the effects of variable skylines, we calculated the median flux value at each spectral channel and subtracted it for each spaxel. This subtraction sets the median flux value to zero for all spectral channels.
    \item \textbf{Flux calibration and telluric correction.} We applied telluric corrections using the telluric standard observation taken closest in time to the science exposures. 
    We use the 2MASS photometric magnitude of the tip tilt star to flux calibrate the data.  First, we calculate the correction factor needed to reproduce the 2MASS magnitude from the observed spectrum of the tip tilt star.  We then apply this correction factor to the spectra in each science frame. The uncertainty in the absolute flux calibration of LGSAO observations is estimated at $\sim 30\,\%$ (see ~\citealt{law09} for further discussion).
    \item \textbf{Mosaicking.} We used the OSIRIS module {\tt Mosaic Frames} to combine all of the science frames into a single spectral cube, using the peak emission of the quasar to align the individual frames for mosaicking. The frames were combined using the sigma-clipping average routine {\tt meanclip2}. Because of the A-B observing pattern and subtraction scheme, this final spectral cube contains a positive central region with negative regions to either side. The positive central region is considered the science region, and only positive spaxels were used for the final analysis. The final spectral cube is shown in Figure~\ref{fig:cube} summed over a $\sim$ 5 $\AA$ wavelength regime centered on \oiiia\ emission. Figure~\ref{fig:cube} visually highlights the background subtraction scheme and the resulting science region.
    \item \textbf{Oversampling.} 
    As done in ~\cite{law07,law09}, we oversampled the spectral cube in order to increase the signal-to-noise ratio in each spaxel, so that we may detect faint, extended emission. We spatially resampled the spectral cube to \oversample \arcsec\ per pixel, and then smoothed with a Gaussian kernel with FWHM=\kernelFWHM\ pixels. The smoothing FWHM is approximately equal to that of the quasar's PSF in the oversampled cube, \fwhmpsf\arcsec.
    \item \textbf{Normalization.} We normalized the continuum in the spectral cube to zero by first choosing a broad noise region on either side of the \oiiia\ emission line, then finding the median continuum value in those two regions for each spaxel. The median continuum value was then subtracted from each spaxel.
  \item \textbf{Wavelength solution.} We checked the wavelength solution using a representative summed spectrum of a background sky region of the cube to ensure that known skylines were consistent with those observed. No correction was necessary.
    \item \textbf{Heliocentric correction.} We corrected the spectra for the heliocentric motion of the earth using the IRAF package {\tt rvcorrect}. All wavelengths and redshifts are reported in the heliocentric vacuum frame, with redshifts being determined using the rest-frame vacuum wavelengths of the emission lines.
    \item \textbf{Determination of spectral resolution.} We estimated the spectral resolution by measuring the FWHM of OH-skylines that are near in wavelength to the redshifted \oiii\ emission lines. We measured the average FWHM of skylines to be $\sim$\sigins\ \kms, very close to the expected average resolution of 100 km s$^{-1}$.  Reported velocity dispersions have been corrected for this instrumental FWHM by subtracting it in quadrature.
\end{enumerate}

  \section{Analysis}\label{sec:analysis}
 In this section, we describe the analysis of the final spectral cube.
 
  \subsection{\oiii\ Flux and Luminosity}
  
\begin{figure*}
\includegraphics[scale=0.4]{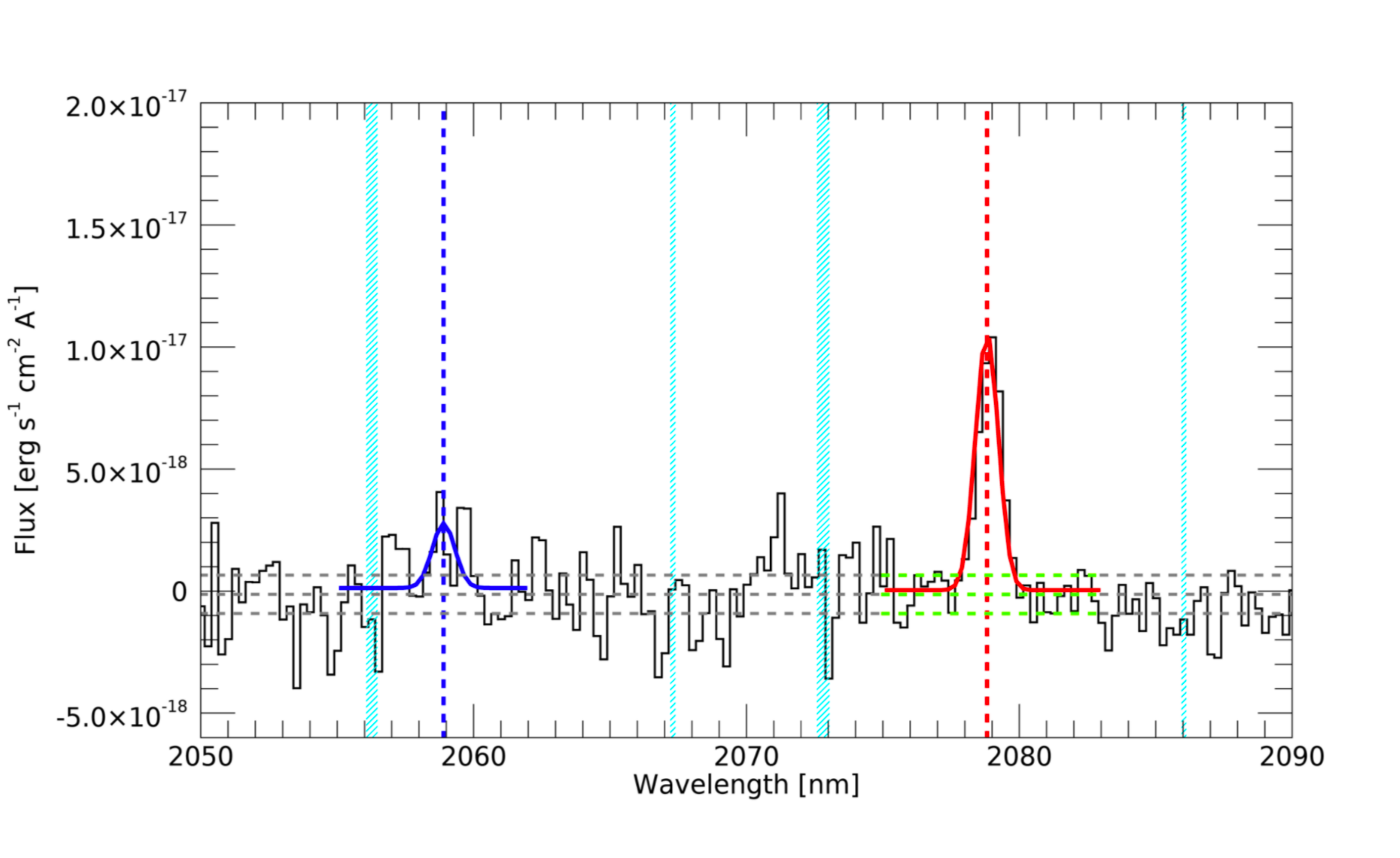}
\caption{Composite spectrum over a circle of of radius 0.3\arcsec\ centered on the peak of \oiiia\ emission.  The \oiiia\ emission is shown with a best-fit Gaussian overlaid in red, while the best-fit Gaussian to the \oiiib\ emission is overlaid in blue.  The best-fit redshift of the \oiiia\ emission is $z = $ $\oiiiazabs$ , indicated here by the vertical red dashed line. We fix the redshift and line width of \oiiib\ to that of \oiiia . Cyan hashed lines indicate strong atmospheric OH emission features, which can leave residuals. The mean noise in the spectrum, $\pm$ 1$\sigma$, is shown with horizontal dashed gray lines. In the region of the spectrum used for noise calculations, these lines are overlaid with green.
}
\label{fig:oiiia}
\end{figure*}

Given the extended nature of the \oiiia\ emission, we measure the total \oiiia\ flux in two ways.  First, we create a composite spectrum that maximizes the S/N of the detected emission.  This composite spectrum, shown in Figure~\ref{fig:oiiia}, was created by summing the spaxels over a circle of radius 0.3\arcsec\ centered on the bright central core of \oiiia\ emission, omitting faint and extended emission. From this `S/N-maximizing' spectrum we report best-fit redshift, line width, and S/N, using a best-fit Gaussian and report the results in Table ~\ref{tab:summarylines}. The uncertainties are calculated by performing a Monte Carlo analysis, in which we perturb the original spectrum using values drawn from a uniform distribution based on the measured noise, and then re-calculate the Gaussian fit. We repeat this procedure 10,000 times and consider the standard deviation of the resulting line centroids and line widths to be the uncertainty in the measurements.  The noise is taken to be the standard deviation of the residual, after subtracting the Gaussian fit, in $\pm$ 5\AA\ around the wavelength of peak emission (region shown in green in Figure ~\ref{fig:oiiia}). 

Second, we create a composite spectrum that includes all of the detectable, faint, extended emission, in order to report the maximum total detected flux (albeit with a lower S/N due to the inclusion of increasingly noisy, faint emission regions). This spectrum, summed over a $1.2 \times 0.8\arcsec $ region, includes all observed emission while carefully excluding the negative of the quasar emission.  We use this second, `flux-maximizing' composite spectrum to report the total detected \oiiia\ flux, F$^{\oiii \lambda5007}$ = $\oiiiaflux$ \flux\, which corresponds to luminosity L$^{\oiii \lambda5007}$ = $\oiiialum$.

Finally, we measure an upper limit on the flux of unresolved \oiiib, by fixing the redshift and linewidth to that of \oiiia.  We report this upper limit in Table~\ref{tab:summaryresults}, but note that because the \oiiib\ line falls near a strong atmospheric OH emission feature, residuals have increased the noise in this spectral region.  Given the canonical ratio of \oiiia/\oiiib\ $\sim$ 3, we expect \oiiib\ flux $\sim$ 8.0$\times$10$^{-18}$ \flux, consistent with the measured limit.

    \subsection{\hbeta\ Flux and Luminosity}
\begin{figure*}

\includegraphics[scale=0.3]{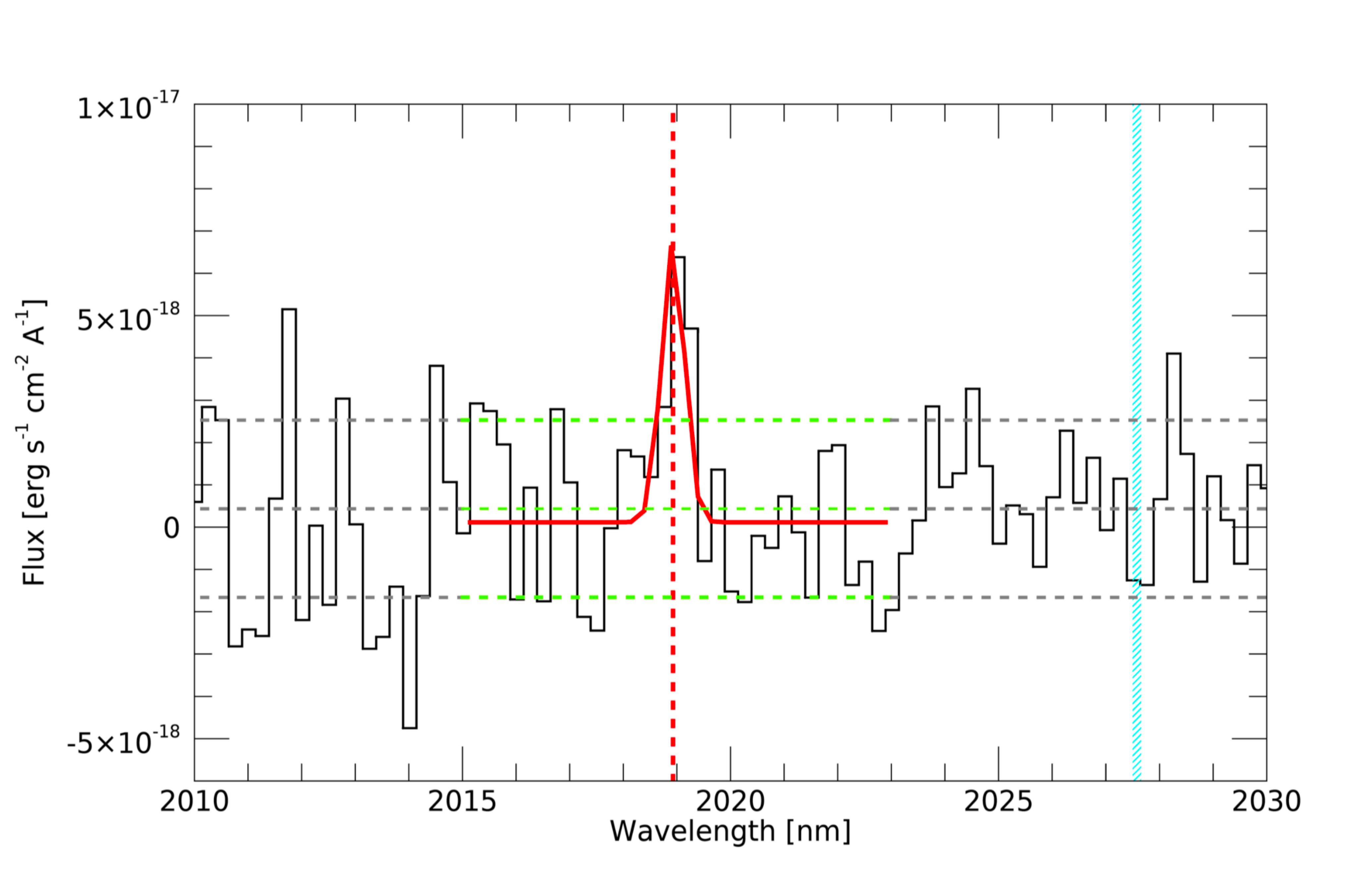}
\caption{Composite spectrum over a 0.3$\times$0.3\arcsec\ region centered on the peak of \hb\ emission. The \hbeta\ emission is shown with a best-fit Gaussian overlaid in red. The best-fit redshift of the \hbeta\ emission is $z = $ \hbetazabs, indicated here by the vertical red dashed line. Cyan hashed lines indicate strong atmospheric OH emission features, which can leave residuals. The mean noise in the spectrum, $\pm$ 1$\sigma$, is shown with horizontal dashed gray lines. In the region of the spectrum used for noise calculations, these lines are overlaid with green.
}
\label{fig:hbeta}
\end{figure*}

We follow the same procedure for \hbeta\ as for the \oiii\ lines, creating a `S/N-maximizing' composite spectrum by summing over a $0.3\times0.3$\arcsec\ square, centered on the same approximate location as the peak of the \oiiia\ emission. The resultant spectrum is shown in Figure \ref{fig:hbeta}. Based on a Gaussian fit to the emission line in this composite spectrum, we report a best-fit redshift $z= \hbetazabsu$ and measure S/N=\hbetasig.

We use a flux-maximizing spectrum summed over a 1.0$\times$0.75\arcsec\ rectangle to measure the flux of \hbeta\ emission. We measure F$^{H\beta \lambda4861}$ = $\hbetaflux$ \flux, which corresponds to L$^{H\beta \lambda4861}$ = $\hbetalum$ \lum\ (not corrected for dust).  Adopted line diagnostics and flux measurements of \hbeta\ emission are summarized in Tables~\ref{tab:summarylines} and \ref{tab:summaryresults} respectively.
       

\LongTables
\begin{center}
\begin{deluxetable}{llc}
\tablewidth{0pc}
\tablecaption{Line Diagnostics of DLA 2233$-$13 \label{tab:summarylines}}
\tabletypesize{\scriptsize}
\tablehead{\colhead{Quantity} & \colhead{Units} &
\colhead{ Measured}} \\\
\startdata
$z_{HIRES}$&...& \hireszabs\ \\
\hline
$z$(\oiii\ $\lambda$5007)$^{a}$ &...& $\oiiiazabsu$\ \\
$\Delta v$(\oiii$\lambda$5007 )$^{a,c}$ & [km s$^{-1}$] & \oiiiavel\ \\
FWHM(\oiii\ $\lambda$5007)$^{a}$ & [km s$^{-1}$] &$\oiiiafwhminit$ \\
FWHM(\oiii\ $\lambda$5007)$^{a,b}$ & [km s$^{-1}$] & $\oiiiafwhm$ \\
\hline

$z$(H$\beta$) && $\hbetazabsu$\ \\
$\Delta v$(H$\beta$)$^{c}$ & [km s$^{-1}$] & \hbetavel\ \\
FWHM(H$\beta$)& [km s$^{-1}$] & $\hbetafwhminit$ \\
FWHM(H$\beta$)$^{b,d}$ & [km s$^{-1}$] & -\ \\
\enddata
\tablenotetext{a}{The \oiiib\ line is taken to have the same redshift and FWHM as the \oiiia\ line.}
\tablenotetext{b}{Effects of instrumental smoothing taken out in quadrature.}
\tablenotetext{c}{Velocity difference between $z_{HIRES}$ and the given transition.}
\tablenotetext{d}{\hb\ emission is unresolved after instrumental smoothing.}
\end{deluxetable}
\end{center}

 \subsection{Star Formation Rate Estimation}
The star formation rate (SFR) is typically estimated using \ha\ emission, as in the ~\cite{kennicutt98schmidt} calibration that relates L$^{H\alpha}$ to star formation rate,

\begin{equation}
\mathrm{SFR (M_{\odot} \ year^{-1})} = 7.9\times10^{-42}\times L^{H\alpha} (\rm{ergs\ s}^{-1})
\label{eqn:sfr}
\end{equation}

Unfortunately, for \mydla\ the \ha\ emission is redshifted out of the Keck/OSIRIS filter range.  Therefore, we estimate the star formation rate using \oiii\ or \hb\ emission. There are three ways that we can estimate the star formation rate:

\begin{enumerate}
    \item Adopt the standard Balmer ratio, \ha/\hb\ = 2.8, to convert L$^{H\beta}$ to L$^{H\alpha}$, and then proceed with the ~\cite{kennicutt98schmidt} relation. With this method, we find a star formation rate of $\sfra$ \sfr.
    \item Follow the method outlined in ~\cite{suzuki15} and assume that \oiii/\ha\ $\sim$ 2.4 (the maximum value for local star-forming galaxies), then proceed with the ~\cite{kennicutt98schmidt} relation to estimate a lower limit star formation rate. With this method, we find that the lower limit of the star formation rate is $\sim \sfrb$ \sfr.
    \item Follow the method outlined in ~\cite{weatherley05} and assume that \oiii/\ha\ $\sim$ 1.67 for \mydla, then proceed with the \cite{kennicutt98schmidt} relation. With this method, we find a star formation rate of $\sim \sfrc$ \sfr.
\end{enumerate}

Each of these methods comes with their own set of assumptions and large inherent uncertainties. However, we  note that the three methods provide estimates that agree within errors (except the ~\cite{suzuki15} method, which is a lower limit).  There is no clear argument for which method is the most accurate. ~\cite{kennicutt92} found a large scatter between L$^{H\alpha}$ and L$^{\oiii}$ for a sample of nearby galaxies, and recommended against using \oiii\ emission as an indicator of star formation rate at all for this reason. On the other hand, ~\cite{weatherley05} suggest that differences in metallicity and ionization parameter within the sample of galaxies could account for this scatter. Further, ~\cite{weatherley05} calculate \oiii/\ha\ as a function of these parameters using relations from ~\cite{kewley02}, and then estimate \oiii/\ha\ for \mydla\ by using its known metallicity and assuming ionization parameters similar to Lyman break galaxies. We have estimated the SFR of \mydla\ with each of these methods and present a comparison to other measurements in 
Section~\ref{sec:results}.

We note that the ratio of \halpha/\hb, used in the first method of calculating a star formation rate, is sensitive to dust extinction. However, as DLAs are relatively low-metallicity systems, with typical metallicities $\sim$ 1/30$^{th}$ solar \citep{rafelski12}, they are generally considered to be unaffected by dust. Furthermore, \cite{rafelski12} found that the [$\alpha$/Fe] ratio of DLAs is relatively constant at low metallicity, noting the absence of expected effects from dust depletion in their sample. While \mydla\ has a metallicity of $\sim$ 1/10$^{th}$ solar (see section~\ref{subsec:met}), and is therefore more metal-rich than the typical DLA, it is still relatively metal-poor with respect to solar, and therefore we consider the assumption \halpha/\hb = 2.8 to be reasonable.

  \subsection{Spatial mapping of intensity, velocity, velocity dispersion, and signal-to-noise ratio}
  
\begin{figure*}
\includegraphics[scale=0.7,trim=100 0 0 0]{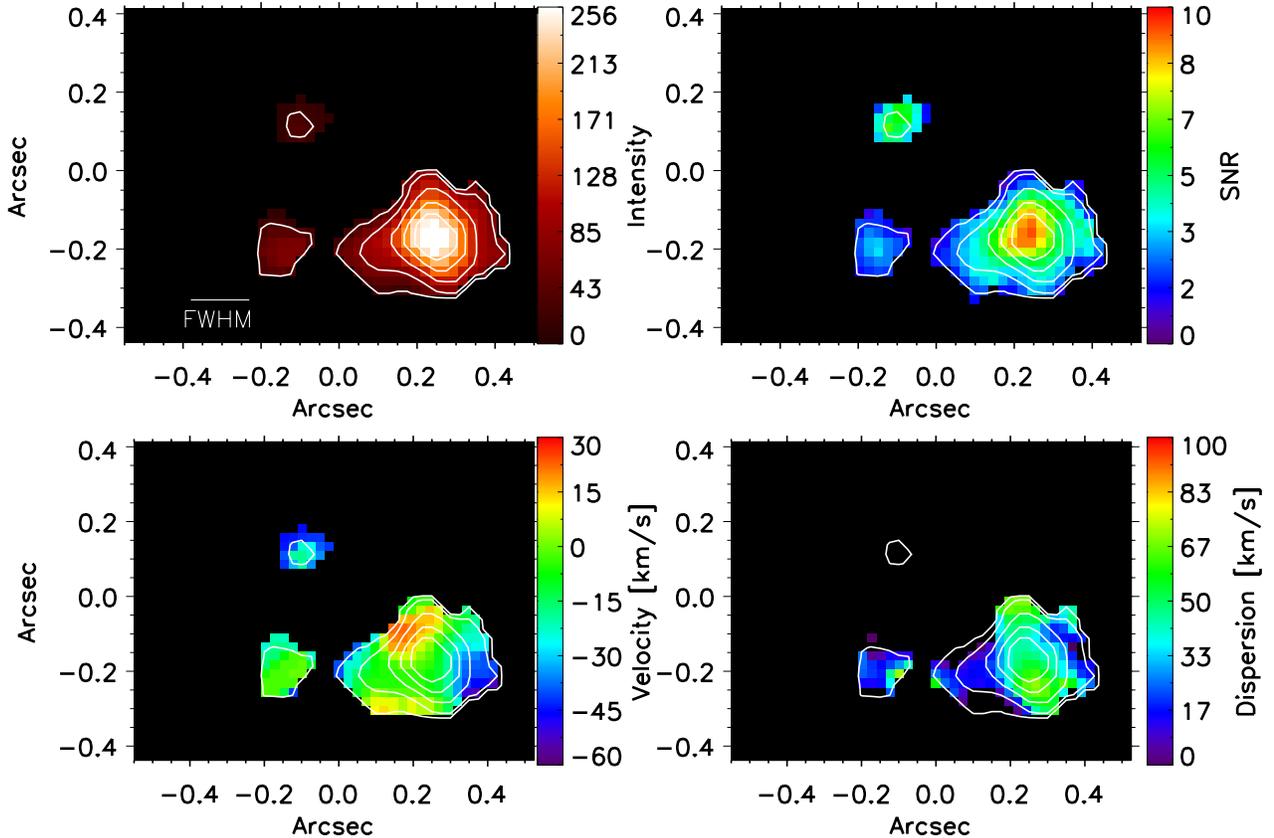}
\caption{ \oiii\ $\lambda$5007 intensity (top left), velocity (bottom left), velocity dispersion (bottom right) and  S/N (top right) maps.   The orientation is the standard North, up and East, to the left.  The velocity is relative to $z =$ \oiiiazabs , the best-fit redshift determined from the composite spectrum shown in Figure~\ref{fig:oiiia}.  Individual spaxels are 0.0025 square arcseconds. The FWHM = \fwhmpsf \arcsec\ of the PSF after smoothing is shown in the lower left corner of the intensity map.  At the redshift of the \dla\ 1\arcsec\ corresponds to \kpcscale\ kpc. The quasar, not shown, is located \quasarsep \arcsec to the left of the emission region. 
}
\label{fig:oiiivelocity}
\end{figure*}

\begin{figure*}
\includegraphics[scale=0.70]{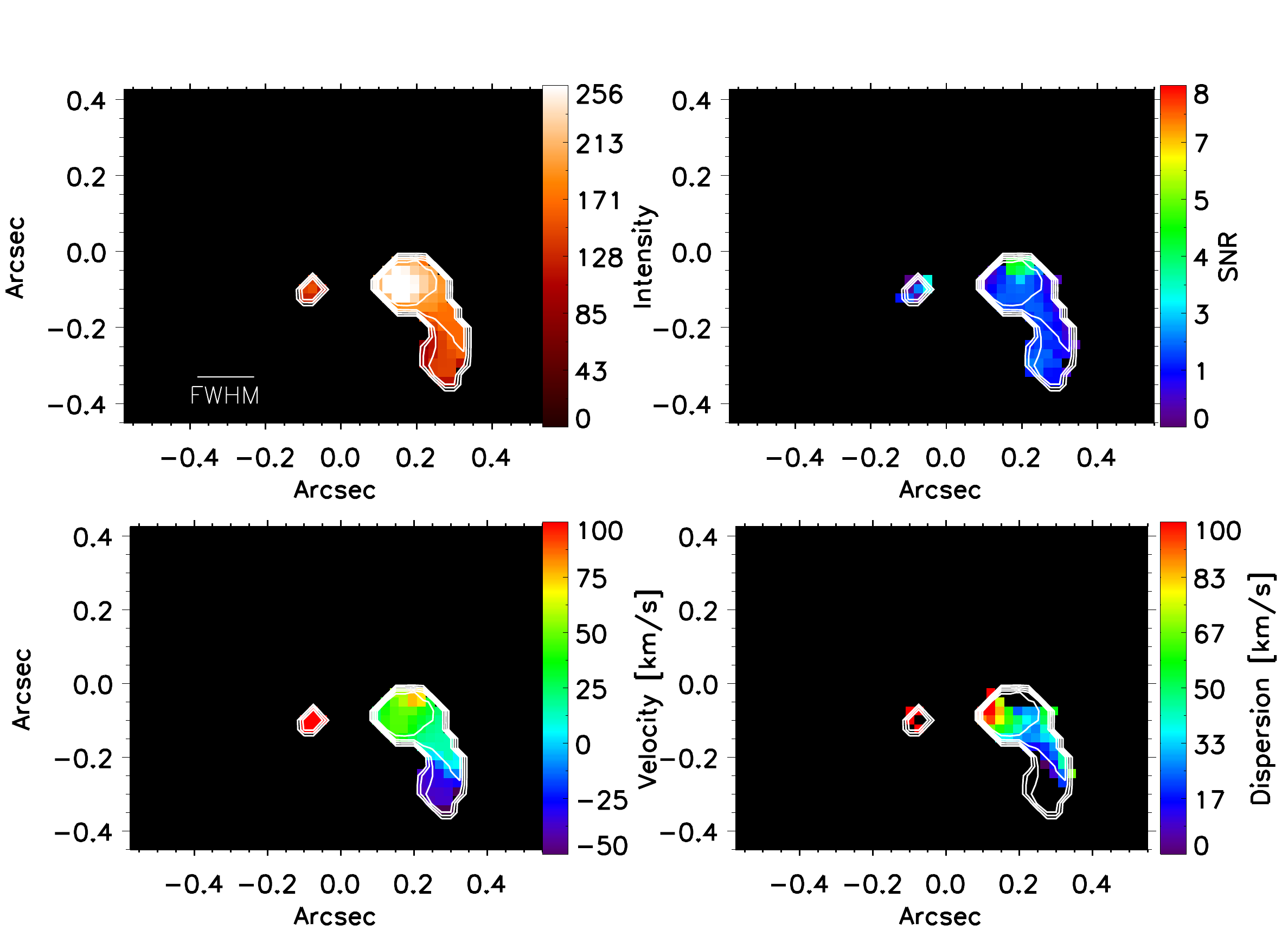}
\caption{\hbeta\ intensity (top left), velocity (bottom left), velocity dispersion (bottom right) and  S/N (top right) maps.   The orientation is the standard North, up and East, to the left.  The velocity is relative to $z =$\hbetazabs , the best-fit redshift determined from the composite spectrum shown in Figure~\ref{fig:hbeta}.  Individual spaxels are 0.0025 square arcseconds. The FWHM = \fwhmpsf \arcsec\ of the PSF after smoothing is shown in the lower left corner of the intensity map.  At the redshift of the \dla\ 1\arcsec\ corresponds to \kpcscale kpc. The quasar, not shown, is located \quasarsep \arcsec to the left of the emission region. 
}
\label{fig:hbetavelocity}
\end{figure*}

 To map the emission and search for kinematic signatures, we fit a Gaussian to the expected location of emission in each individual spaxel in the science region of the data cube and compared the chi-squared result to that of a fit with no emission line. We require a minimum of 1$\sigma$ for detection. Additionally, we perform a visual inspection of the spectrum in each spaxel and reject spaxels where the spectrum appears to be dominated by noise features or spikes, or where the emission features are not well fit by a Gaussian. If no line is detected or any quality check is failed, we do not report a detection and the spaxel is black in the emission maps. All spaxels where we find a credible detection are displayed in Figure~\ref{fig:oiiivelocity} and Figure~\ref{fig:hbetavelocity}.
 
 For spaxels where emission is detected, we calculate a best-fit flux, velocity (relative to the best-fit redshift determined from the composite spectrum of the emission region), velocity dispersion, and signal to noise ratio. Velocity dispersion values have been corrected for the instrumental resolution by subtracting instrumental FWHM in quadrature. In this way, we generated two-dimensional maps of the \oiiia\ and \hbeta\ emission, which we present in Figure~\ref{fig:oiiivelocity} and Figure~\ref{fig:hbetavelocity}. These figures show an intensity map (top left), velocity map (bottom left), velocity dispersion map (bottom right), and S/N ratio (top right) in \oiiia\ and \hbeta\ respectively. The average FWHM of the point-spread function (PSF) is \fwhmpsf\arcsec, indicated by the white bar. The FWHM was measured from the image of the quasar. The quasar is \quasarsep\arcsec\ from the center of \dla\ emission, which corresponds to a physical distance of \quasardist\ kpc. The location of the quasar is outside the field of view of the maps and is not shown.
 
We measure the S/N of the line detection in each spaxel by subtracting the Gaussian fit from the spectrum and taking the standard deviation of the residual to be the noise in that spaxel. The S/N ratio is the ratio of the amplitude of the best-fit Gaussian in that spaxel to the measured noise. 

In some spaxels, the instrumental resolution is greater than the FWHM of the Gaussian fit to the emission line, resulting in a nonreal value when we remove instrumental resolution in quadrature.  Emission in these spaxels is unresolved. We are able to measure intensity, velocity, and S/N ratio, but not velocity dispersion. For this reason, these spaxels are black in the velocity dispersion maps.

\section{Results}~\label{sec:results}

In this section, we place these measurements in the context of previous observations of \mydla\ and other high-z DLAs and sub-DLAs. We summarize these current results in Table~\ref{tab:summaryresults}.

\LongTables
\begin{center}
\begin{deluxetable}{llc}
\tablewidth{0pc}
\tablecaption{Summary of Results for DLA 2233$-$13 \label{tab:summaryresults}}
\tabletypesize{\scriptsize}
\tablehead{\colhead{Quantity} & \colhead{Units} &
\colhead{ Measured}} \\
\startdata
F(\oiiia)$^{a}$ & [ergs s$^{-1}$ cm$^{-2}$] & $\oiiiaflux$   \\
L(\oiiia)$^{a}$ & [ergs s$^{-1}$]   & $\oiiialum$ \\
SFR(\oiiia)$^{b}$ & [M$_{\odot}$yr$^{-1}$] & $\sfrc$ \\

F(\oiiib)$^{a}$ &  [ergs s$^{-1}$ cm$^{-2}$] & $\oiiibflux$ \\
L(\oiiib)$^{a}$ & [ergs s$^{-1}$] & $\oiiiblum$ \\

F(\hbeta) & [ergs s$^{-1}$ cm$^{-2}$] & $\hbetaflux$   \\
L(\hbeta ) & [ergs s$^{-1}$]   & $\hbetalum$ \\
SFR(\hbeta)$^{c}$ & [M$_{\odot}$yr$^{-1}$] & $\sfra$ \\ 
M$_{dyn}$ & [M$_{\odot}$] & $\oiiiamass$ \\
M$_{gas}$ & [M$_{\odot}$]& $\totgas$ \\
$f_{gas}$ & - &$\sim 60\%$ \\
\enddata
\tablenotetext{a}{1$\sigma$ uncertainties are determined after subtraction of the Gaussian model.}
\tablenotetext{b}{SFR(\oiii) calculated by assuming \ha/\oiiia $\sim$ 0.6 ~\citep{weatherley05} and then using the empirical relationship between L(\ha) and SFR in \cite{kennicutt98schmidt}.}
\tablenotetext{c}{SFR(\hb) calculated by assuming a standard Balmer ratio and then using the empirical relationship between L(\ha) and SFR in \cite{kennicutt98schmidt}.}
\end{deluxetable}
\end{center}

\begin{figure}
\includegraphics[scale=0.5]{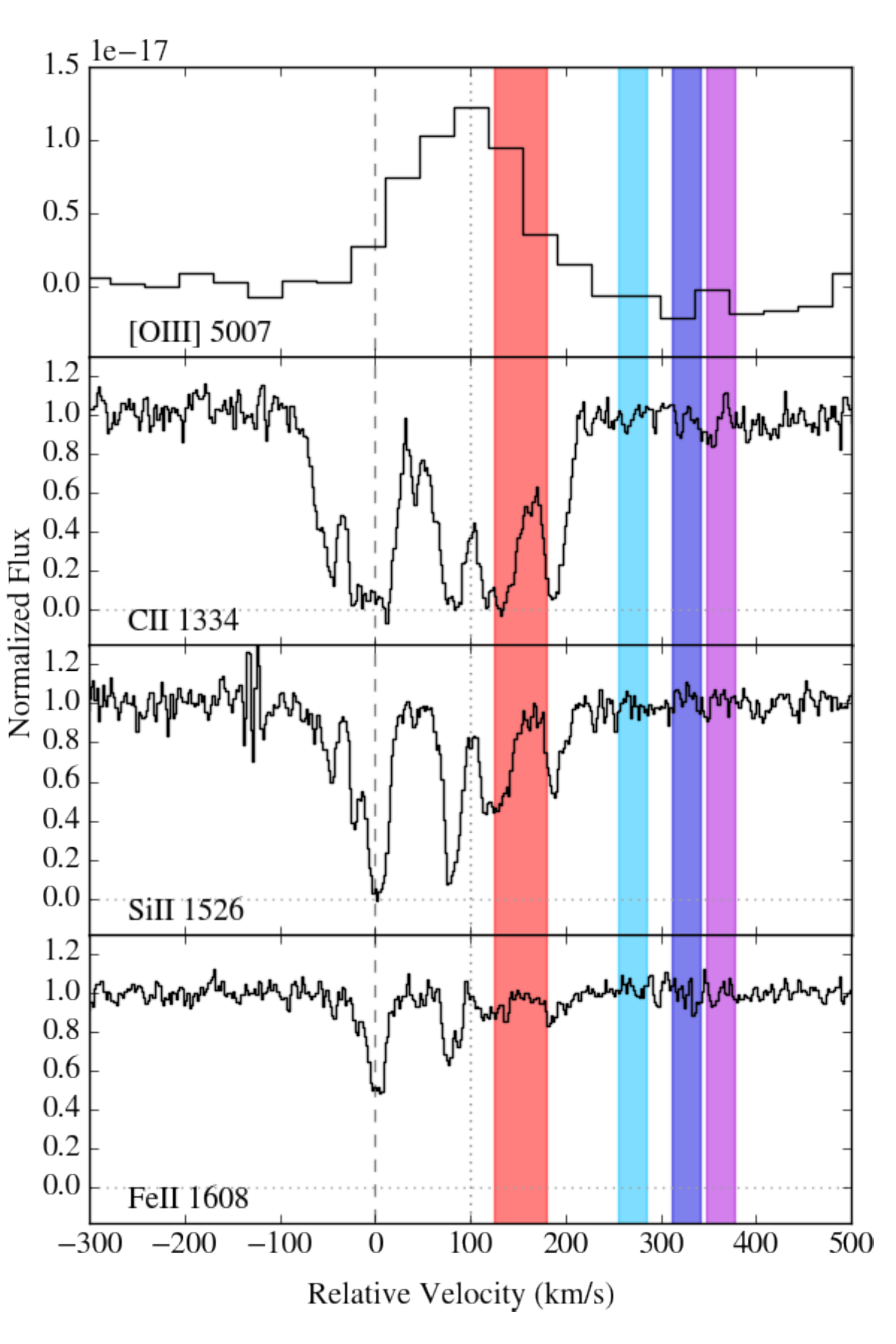}
\caption{The composite \oiiia\ spectrum of \mydla\ (top panel) shown with representative low-ion transitions from the Keck/HIRES spectrum (bottom three panels).  The grey dashed line denotes the adopted absorption redshift \hireszabs , while the grey dotted line denotes the best-fit \oiiia\ emission redshift, \oiiiazabs .  For reference, the red bar indicates the velocity centroid and FWHM of the previous [OIII] detection ~\citep{weatherley05}, while the cyan, blue, and violet bars indicate the the offsets of previous Lyman-$\alpha$ emission detections (respectively, ~\cite{djorgovski96, christensen04, christensen07}) as summarized in Table~\ref{tab:previous}.  Given the large FWHMs of the Lyman-$\alpha$ we have indicated them here with 30 \kms-wide bars for clarity.  
}
\label{fig:abs}
\end{figure}

\subsection{Metallicity}~\label{subsec:met}

In the literature, \mydla\ is most commonly referred to as having a metallicity of [Fe/H]\footnote{We use the standard shorthand notation for metallicity relative to solar, [M/H] = log(M/H) $-$ log(M/H)$_{\odot }$.} = -1.4 based on ~\cite{lu1997} estimation from their Keck/HIRES spectrum.  We reanalyzed the Keck/HIRES spectrum and followed the procedure of ~\cite{rafelski12} for estimating DLA metallicities.  Unfortunately, transitions of the preferred ions of O, S, Si and Zn are either saturated, heavily blended, or not covered in the Keck/HIRES spectrum, leaving us to estimate the metallicity from the unsaturated FeII transition (see Figure~\ref{fig:abs}).  We use the AODM technique ~\citep{savage91} and measure [Fe/H] = $-1.36 \pm 0.02$.  We then apply the ~\cite{rafelski12} $\alpha$-enhancement correction to find [M/H] = [Fe/H] $+$ 0.3 dex = $-1.06$, confirming that this is a relatively metal-rich system, given the typical DLA metallicity of [M/H] $\sim$ $-1.5$ ~\citep{rafelski12}. We note that the potential depletion of Fe, which is larger in higher metallicity systems, may mean that this estimation is, strictly speaking, a lower limit on the metallicity of this system.

\LongTables
\begin{center}
\begin{deluxetable}{llccc}
\tablewidth{0pc}
\tablecaption{Previous and Current Emission-Line Measurements of \mydla\ \label{tab:previous}}
\tabletypesize{\scriptsize}
\tablehead{\colhead{Line} & \colhead{Flux } & \colhead{FWHM} & \colhead{Redshift} & \colhead{Reference}\\
 & \colhead{[ergs s$^{-1}$ cm$^{-2}$] } & \colhead{[km s$^{-1}$]} &  }\\
\startdata
\lya & (6.4 $\pm$ 1.2)$\times$10$^{-17}$ & 200 & 3.1530 & a  \\
\lya & (2.8 $\pm$ 0.3)$\times$10$^{-16}$ & 1090 & 3.1538 & b  \\
{\lya} & (9.6 $\pm$ 2.5)$\times$10$^{-17}$ & 230 &   3.1543 & c \\
\oiiia & (6.78 $\pm$ 0.5)$\times10^{-17}$ &  55 & 3.15137 & d \\
\oiiia & $\oiiiaflux$ & 84 &   $\oiiiazabs$ & e \\
\oiiib & $\oiiibflux$ & 85 & $\oiiibzabs$ & e \\
\hbeta & $\hbetaflux$ & f & $\hbetazabs$ & e \\
\enddata
\tablenotetext{a}{~\cite{djorgovski96}, Keck/LRIS observations. 
}
\tablenotetext{b}{~\cite{christensen04}, PMAS IFU observations, summed over 3 by 5 arcsec region and corrected for Galactic reddening.} 
\tablenotetext{c}{~\cite{christensen07}, higher spectral resolution PMAS IFU observations.}
\tablenotetext{d}{~\cite{weatherley05}, VLT/ISAAC observations.}
\tablenotetext{e}{This work}
\tablenotetext{f}{Unresolved line}
\end{deluxetable}
\end{center}

\subsection{Discrepancy with previous flux measurements}~\label{subsec:flux}

We note that the flux we measure is not in agreement with that of \cite{weatherley05}, who measured \oiiia\ flux of $6.78\pm0.5 \times 10^{-17}$ \flux\ (compare to our measurement, F$^{\oiii \lambda5007}$ = $\oiiiaflux$ \flux). This discrepancy may be due to the location of the \dla\ emission. The emission region is very close to the edge of the field of view. It is possible that the observations presented here do not capture the full extent of emission, resulting in a flux measurement that is lower than that of ~\citet{weatherley05}. 

Given this discrepancy, we perform two additional tests of the fidelity of the flux estimation.  First, we model the potential loss of flux due to the \dla\ emission being close to the edge of the science region.  In this model, we assume that there is a faint, extended "arm" of emission that mirrors the `arm' we observe to the left of the central core of emission. We measure the flux in the summed spectrum of the left `arm' to be (6.4 $\pm$ 2.3)$\times10^{-18}$ \flux. This summed spectrum is taken over a 0.75$\times$1.2\arcsec\ area directly to the left of and just excluding the central core of the emission region. We assume that an equal amount of flux may be located outside the field of view to the right of the DLA and add it to the original flux measurement. This results in an estimated total flux of (3.0 $\pm$ 0.55)$\times10^{-17}$ \flux, with a 1-$\sigma$ upper limit, accounting for the additional $\sim30$\% flux calibration uncertainty, of $4.7\times10^{-17}$ \flux. This adjusted flux estimate is in better agreement with that of ~\cite{weatherley05}.

Second, we verify the flux calibration by calculating the flux of the observed standard star using the same procedure we apply to the DLA emission (see \S2.1, bullet point 3).  As a result, we reproduce the 2MASS magnitude of the standard star to within 10\%.

\subsection{Star formation rate}
We estimate the star formation rate in three different ways using \oiii\ and \hbeta\ luminosities: one by using the standard Balmer ratio to convert \hb\ to \ha\ luminosity, and two by assuming a value of \oiii\ /\ha\ (described in detail in \S3.3). We find SFR values ranging from 7.1 \sfr\ to 13.6 \sfr.

There are two previous estimates of SFR for this galaxy. With the method outlined above, \cite{weatherley05} find that \mydla\ has a star formation rate of 28 \sfr, and suggest that this method is accurate to within a factor of 2. The discrepancy between the \citet{weatherley05} measurements and those presented here stems from a difference in measured flux as discussed in \S4.1.

\cite{christensen04} also provides an estimate of the SFR, but based on measured \lya\ flux and the assumption that \lya/\ha\ $\sim$ 10. This assumption makes it possible to convert \lya\ flux to \ha\ flux and calculate the SFR based on the \cite{kennicutt98schmidt} relation. \cite{christensen04} measure a SFR of $\sim$ 19 $\pm$ 10 \sfr. The estimates presented here (and the lower limit based on \citealt{suzuki15}) are in agreement with this measurement. 
 
There are some estimates of star formation rates for other DLAs in the literature. For redshift $2\leq z\leq 3$, DLAs have star formation rates spanning $2$ $M_{\odot}$ to 17 $M_{\odot}$ ~\citep{jorgenson14,krogager13,fynbo10,peroux12,srianand2016}. The measurements presented here are well within the range of previously measured SFR for DLAs.

\subsection{Dynamical mass estimate}
We estimate the dynamical mass within the radius of \oiiia\ emission using equation 2 from ~\cite{law09},
  
   \begin{equation}
  M_{dyn} = \frac{C \sigma ^{2}_{net} r}{G}
   \end{equation}
  
\noindent   where $C = 5$ for a uniform sphere ~\citep{erb06}, $\sigma_{net}$ = \oiiiasignet\ \kms, and $r$ is the radius of emission. $\sigma_{net}$ is an overall velocity dispersion, measured from the Gaussian fit to the \oiiia\ composite spectrum (see \S3.1 and Figure~\ref{fig:oiiia} for a description of the composite spectrum). 
We estimate visually that the radius of emission for this DLA is 0.2\arcsec, or 1.5 kpc.
We use \oiiia\ emission to estimate the emission radius because it is the strongest line we observe. Estimating the radius of emission using the less-extended \oiiib\ or \hbeta\ emission would be likely to under-estimate the dynamical mass.  Based on this estimate of emission radius, we find dynamical mass $M_{dyn} =$ $\oiiiamass$ M$_{\odot}$.


There are few estimates of dynamical mass for DLAs and sub-DLAs in the literature: \cite{srianand2016} found $M_{dyn}\sim (1-6)\times10^9$ $M_{\odot}$ for a DLA at $z=2.9791$, \cite{jorgenson14} and \cite{krogager13} measured $M_{dyn}\sim 6.1\times 10^9$ $M_{\odot}$ and $M_{dyn} \sim 2.5\times10^9$ $M_{\odot}$ respectively for the same $z=2.35$ DLA, and \cite{bouche2013} found $M_{dyn}\sim 2\times10^{10}$ $M_{\odot}$ for a $z\sim 2.3$ DLA. These estimates for \dla\ galaxies, including \mydla\ presented here, are generally consistent with the dynamical mass of star-forming galaxies at high redshift, estimated to be $3\times 10^9 M_{\odot} \leq M_{dyn} \leq 25 \times 10^9 M_{\odot}$ by \cite{law09}. The dynamical mass of \mydla\ falls into the low end of this range.

At a lower redshift range, $0\leq z \leq 1$, DLA and sub-DLAs have similar dynamical mass ranges, spanning $7.9\times10^9$ $M_{\odot}$ to $7.9\times10^{10}$ $M_{\odot}$ ~\citep{peroux16,peroux11,chengalur02}.

  
  \subsection{Gas mass estimates}
   We estimate the gas mass of the galaxy using the Kennicutt-Schmidt relation ~\citep{kennicutt98schmidt}. We first estimate the star formation rate surface density by dividing star formation rate by the area over which we detect \oiiia\ emission. We can then use equation 4 from ~\citet{kennicutt98schmidt} to convert $\Sigma_{SFR}$ to $\Sigma_{gas}$, 
   
   \begin{equation}
       \Sigma_{gas} = (\Sigma_{SFR})^{\frac{1}{1.4}}\times (2.5\times10^{-4})^{-\frac{1}{1.4}}
   \end{equation}
   
\noindent   and convert the gas mass surface density to a total gas mass by multiplying it by the area of the emission region. For the estimates of star formation rates for \mydla\ detailed in section \S3.4, we find, $\totgasb M_{\odot} \leq M_{gas} \leq \totgasc$ $M_{\odot}$.  This estimate is lower than the average gas mass for $z$ $\sim2$ star-forming galaxies, measured by \cite{erb06} as $\langle M_{gas} \rangle = 2.1\pm 0.1\times 10^{10}$ $M_{\odot}$. It is the same order of magnitude as measurements of gas mass by \cite{jorgenson14} and \cite{krogager13} for the same z $\sim 2.3$ DLA of $M_{gas} = 4.2\times10^9$ $M_{\odot}$ and $1\times 10^9$ $M_{\odot}$ respectively.
   
We estimate the gas fraction of \mydla , \begin{equation}
f_{gas} = \frac{M_{gas}}{M_{gas} + M_{dyn}} 
\end{equation}to be $f_{gas} \simeq$ \gasfrac, also similar to those found in \cite{jorgenson14} and \cite{krogager13}.

\begin{figure*}
\begin{center}\includegraphics[scale=0.65, trim=150 50 150 150]{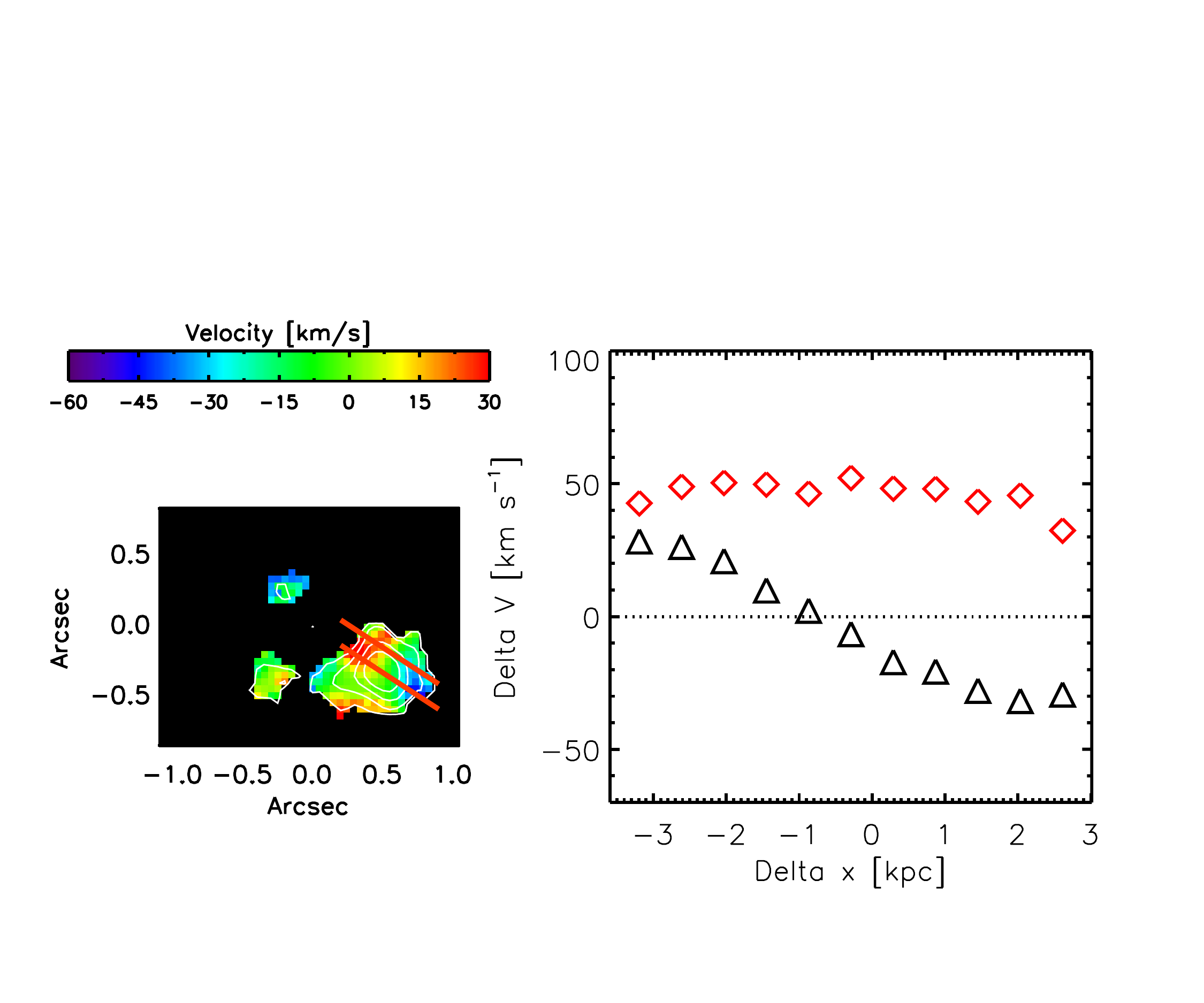}\end{center}
\caption{Artificial slit laid down on \oiiia\ velocity map. The slit is 0.15\arcsec\ wide. Velocity and velocity dispersion are sampled in 0.075\arcsec\ wide regions along the slit and shown in the right panel. Relative velocity (black triangles) and velocity dispersion (red diamonds) are shown in the right panel. The x-axis is position along the slit in kiloparsecs, and the y-axis is the difference between the measured velocity and the velocity of the central redshift.
}
\label{fig:hslit}
\end{figure*}

\subsection{Morphology and kinematics}
 
 \subsubsection{Artificial Slit Analysis}
We use the emission maps shown in Figures~\ref{fig:oiiivelocity} and~\ref{fig:hbetavelocity} to search for signs of kinematic structure. The \oiiia\ emission is more extended and has approximately circular morphology, while the \hbeta\ emission is less extended and slightly elongated in shape, albeit at a lower signal to noise ratio.  While the circular morphology is consistent with a face-on disk, the red- and blue-shifted clumps could indicate disk rotation viewed edge-on. One apparent kinematic axis presents itself. To assess quantitatively whether a rotational signature appears in these maps, we lay down an artificial slit along this kinematic axis, shown in Figure ~\ref{fig:hslit}. The slit width is 0.15\arcsec, matching the LGSAO-corrected seeing, and we sample 0.07\arcsec\ regions to extract velocity and velocity dispersion.

Moving along the slit from the peak of the emission at velocity $v = 0$ \kms\ (indicated by the centroid of the contours), to the upper left is a redshifted clump at $v \sim$ 25 \kms. To the lower right is a blueshifted clump at $v \sim$ -40 \kms. While these features are reminiscent of a rotating disk viewed edge-on, we observe $\Delta v_{rot} \sim$ 65 \kms, which is small compared to $v_{rot}$ for a rotating disk galaxy. Additionally, the observed velocity dispersion is not consistent with this interpretation. In an edge-on rotating disk model, one would expect to see the velocity dispersion peak at the center of emission. As can be seen in Figure~\ref{fig:hslit}, the velocity dispersion (red diamonds) remains high across the emission region.  Even if the velocity dispersion were consistent with a rotating disk, an analysis along a single kinematic axis would be inconclusive at best.

\subsubsection{Kinematic Summary and Analysis}

In Table~\ref{tab:previous} and Figure~\ref{fig:abs} we summarize the emission line measurements of \mydla , including all previous measurements from the literature and those of the current work.

We note when discussing these results that a direct comparison between \lya\ emission and nebular lines like \oiiia\ is not immediately instructive because the \lya\ line is affected by resonant scattering, which affects the shape of the emission line profiles. Rather, we discuss these previous studies to provide a comprehensive overview of this DLA.

Other studies of \mydla\ have also been inconclusive with regard to kinematics. \cite{djorgovski96} observed \mydla\ in \lya\ emission via single slit spectroscopy. The emission was measured 2.3\arcsec\ from the quasar at position angle 159$^{\circ}$, which is consistent with the observations presented here. The authors measured the peak of \lya\ emission at $z = 3.1530\pm0.003$, which corresponds to rest frame $\Delta v= 208$ \kms\ relative to the centroid of the absorption system at $z = $ \hireszabs. While the ~\citet{djorgovski96} measurement of $\Delta v$ is consistent with $v_{rot}$ for a normal disk galaxy, it is insufficient evidence to conclude that the rotating disk interpretation is correct.

Similarly, \cite{christensen04} explored the possibility of disk rotation in \mydla\ with an analysis of composite spectra representing large regions of the 3$\times$5 \arcsec\ field of view. Upon breaking the field of view into four sub-regions and measuring the shifts  between the emission peaks in the east-west and north-south regions of the field, the authors find shifts of 2.5\AA\ ($\sim$ 150 \kms) and 5\AA\ ($\sim$ 300 \kms) respectively. While these velocities may be consistent with $v_{rot}$ for a disk galaxy, more conclusive evidence is needed to support the rotating disk interpretation. ~\cite{christensen04} instead speculate that \mydla\ has multiple emission regions based on the observed double-peaked \lya\ emission. The possible clumpy nature of DLAs is discussed further in ~\cite{augustin18}. 

 To date, no observations of \mydla\, including those presented here, have yielded evidence that is conclusively in favor of an edge-on rotating disk.

\section{Summary}~\label{sec:summary}   
We present Keck/OSIRIS IFU observations of \mydla, a  $z =$ \myzabs\ sub-DLA, or super Lyman limit galaxy. With the use of LGSAO, we are able to marginally resolve extended \oiiia\ emission from \mydla\ with S/N=\oiiiasig. Further, we detect and report unresolved \oiiib\ and \hbeta\ emission with S/N=\oiiibsig\ and \hbetasig\ respectively. We find a dynamical mass $M_{dyn} = (0.24-4.24)\times10^9$ $M_{\odot}$ and estimate a star formation rate using both \oiii\ and \hb\ emission, finding SFR $\sim 6.5-9.4$ \sfr.

With these marginally spatially resolved observations, we are able to map the \oiiia\ and \hbeta\ emission to search for kinematic signatures that may help to determine the nature of the host galaxy. While \mydla\ does display small clumps of red- and blue-shifted emission, which initially seem indicative of possible edge-on disk rotation, a careful analysis shows them to be inconsistent with the interpretation of an edge-on rotating disk. These observations demonstrate, along with those of \citet{jorgenson14}, the use of Keck/OSIRIS + LGSAO for spatially resolving faint DLA emission, particularly the potential for performing a detailed kinematic analysis that is clearly necessary for understanding the true nature of these enigmatic galaxies.

\acknowledgments
The authors would like to thank the anonymous referee for comments that have significantly improved the manuscript.  R. A. J. gratefully acknowledges support from the NSF Astronomy and Astrophysics Postdoctoral Fellowship under award AST-1102683.  R. A. J. and H. C. acknowledge support from NSF-REU grant AST-1358980 as well as the generous support of the Nantucket Maria Mitchell Association.  The authors gratefully acknowledge the support of the Theodore Dunham, Jr. Grant of the Fund for Astrophysical Research. 
The data presented herein were obtained at the W. M. Keck Observatory, which is operated as a scientific partnership among the California Institute of Technology, the University of California and the National Aeronautics and Space Administration. The Observatory was made possible by the generous financial support of the W. M. Keck Foundation.
The authors wish to recognize and acknowledge the very significant cultural role and reverence that the summit of Mauna Kea has always had within the indigenous Hawaiian community.  We are most fortunate to have the opportunity to conduct observations from this mountain.  

\bibliographystyle{apj.bst}
\bibliography{references.bib}

\end{document}